\documentclass[10pt,letterpaper]{article}
\usepackage{opex3}

\usepackage{graphicx}
\usepackage{dcolumn}
\usepackage{bm}
\usepackage{SIunits}
\usepackage{array}
\usepackage{wasysym}
\usepackage{hyperref}
\usepackage{calc}
\usepackage{cite}

\begin{document}

\title{Optimised quantum hacking of superconducting nanowire single-photon detectors}

\author{Michael G. Tanner,\textsuperscript{1,2,*} Vadim Makarov,\textsuperscript{3} and Robert H. Hadfield\textsuperscript{1,2}}
\address{
\textsuperscript{1}School of Engineering, University of Glasgow, Glasgow, G12~8QQ, Scotland, UK\\
\textsuperscript{2}Institute of Photonics and Quantum Sciences, Scottish Universities Physics Alliance and School of Engineering and Physical Sciences, Heriot-Watt University, Edinburgh, EH14~4AS, Scotland, UK\\
\textsuperscript{3}Institute for Quantum Computing, University of Waterloo, Waterloo, Ontario N2L~3G1, Canada
}

\email{\textsuperscript{*}Michael.Tanner@glasgow.ac.uk}



\begin{abstract*}
We explore bright-light control of superconducting nanowire single-photon detectors (SNSPDs) in the shunted configuration (a practical measure to avoid latching). In an experiment, we simulate an illumination pattern the SNSPD would receive in a typical quantum key distribution system under hacking attack. We show that it effectively blinds and controls the SNSPD. The transient blinding illumination lasts for a fraction of a microsecond and produces several deterministic fake clicks during this time. This attack does not lead to elevated timing jitter in the spoofed output pulse, and hence does not introduce significant errors. Five different SNSPD chip designs were tested. We consider possible countermeasures to this attack.\\
\end{abstract*}


\bibliographystyle{osajnl}


\section{\label{sec:introduction}Introduction}

Quantum communication technologies offer information processing power having no analogues in the classical world. For example, quantum key distribution (QKD) \cite{bennett1984} has been commercialised \cite{comqkdsystems}; secret sharing, quantum teleportation \cite{bouwmeester1997}, entanglement swapping, bit commitment, and blind quantum computation \cite{barz2012} have been demonstrated. To achieve quantum communications at high speed and over long distance, single-photon detectors with high timing resolution and low noise are essential. Superconducting nanowire single-photon detectors (SNSPDs) \cite{natarajan2012} achieve the best combination of these parameters at $1550\,\nano\meter$, the optimal wavelength for long distance transmission over optical fiber.

The first proof-of-principle demonstration using SNSPDs in QKD was carried out with a phase encoding system operating the Bennett-Brassard 1984 (BB84) protocol \cite{bennett1984,hadfield2006}.  A high bit rate, short wavelength ($\lambda = 850\,\nano\meter$) demonstration was then reported based on the Bennett 1992 (B92) protocol with polarization encoding \cite{collins2007,bennett1992}. A high bit rate long distance demonstration at $\lambda = 1550\,\nano\meter$ was carried out at Stanford University \cite{takesue2007} using the differential phase shift (DPS) QKD protocol \cite{inoue2002}. This first QKD demonstration in excess of $200\,\kilo\meter$ ($40\,\deci\bel$ transmission loss) was achieved using SNSPDs with 0.7\% efficiency at $1550\,\nano\meter$, $10\,\hertz$ dark count rate and $60\,\pico\second$ full-width at half-magnitude (FWHM) jitter \cite{takesue2007}. Record bit rates were also achieved at shorter distances, which was a significant improvement on the best QKD results achieved at that time with InGaAs single-photon avalanche photodiodes (SPADs) \cite{gobby2004}.

Since that study, many further QKD demonstrations have been reported using SNSPDs:
the maximum range was extended to $250\,\kilo\meter$ using low-loss fiber and implementing the coherent one-way (COW) protocol \cite{stucki2009,stucki2008}, more recently extended to $260\,\kilo\meter$ with DPS-QKD \cite{Wang2012};
decoy-state protocols \cite{lo2005} have been demonstrated \cite{rosenberg2009,liu2010};
entanglement-based QKD has been demonstrated over long distance \cite{Honjo2008};
SNSPDs have been implemented in QKD field trials in installed fiber networks \cite{Tanaka2008,Choi2010,sasaki2011}.
A detailed comparison between SNSPDs and Si SPADs for short haul high bit rate QKD has also been published \cite{clarke2011}. Since this time SNSPD technology has advanced rapidly \cite{natarajan2012} and near-unity efficiency coupled with low dark count rates is now achievable \cite{marsili2012}. These new high-performance SNSPDs have recently been deployed in interfacing quantum networks with quantum memories via teleportation\cite{bussieres2014}, and are likely to be employed in future QKD demonstrations. The 100-fold improvement in detection efficiency as compared to early demonstrations \cite{takesue2007} in principle allows for 100 times more fiber attenuation, making QKD over up to $60\,\deci\bel$ channel loss feasible. While the current record for highest QKD bit rate \cite{dixon2010} has been achieved using SPADs, it is clear that SNSPDs are a vital technology in the advancement of fiber-based QKD systems and networks particularly with the advent of next generation SNSPDs with near-unity efficiency at telecom wavelengths \cite{marsili2012}.

Information security is an intrinsic feature of quantum communication protocols, guaranteed in principle by the underlying laws of physics \cite{wootters1982,scarani2009}. However, the limitations of components lead to vulnerability. Practical attacks breaking security of QKD have been proposed and successfully demonstrated, by exploiting imperfections and behavior of real hardware not accounted for in the theoretical treatment of security. Several of these attacks exploit imperfections of single-photon detectors, which have mostly been demonstrated on SPAD-based detectors \cite{zhao2008,lydersen2010a,lydersen2010b,weier2011,jiang2013.PhysRevA-88-062335,bugge2014.PhysRevLett-112-070503}. It has been shown by Lydersen {\it et al.} \cite{lydersen2011c} that an SNSPD also has exploitable imperfections, allowing bright-light blinding and deterministic control. A Japanese team has recently applied this technique to explore the vulnerability of DPS-QKD systems and test a countermeasure \cite{Fujiwara2013,Honjo2013}; however they have only investigated blinding of the detector but not optimisation and properties of the fake pulses.


\begin{figure}[t]
\centering\includegraphics[width=\columnwidth]{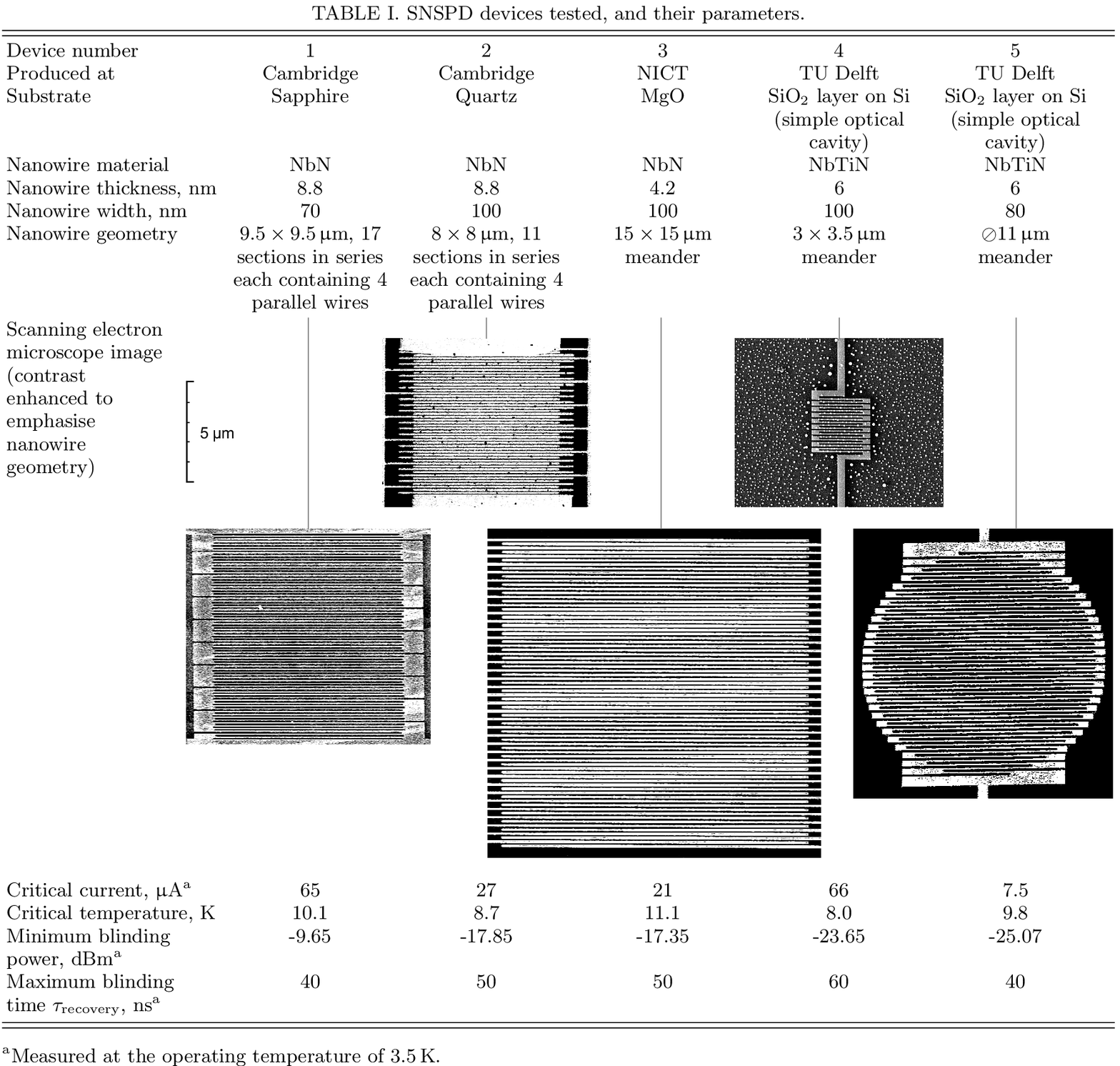}
\label{tab:devices}
\end{figure}

Here we extend the basic technique of detector control by testing and demonstrating this vulnerability in several different SNSPD devices, using a realistic electronic bias and readout that has been employed in QKD demonstrations \cite{hadfield2006}. We demonstrate optimised on-demand fake pulse generation. We also discuss and test non-ideal characteristics of the detector output during the fake pulse generation, and countermeasures to this attack. Although we have only tested stand-alone detectors, our findings reflect on the security of any QKD system that would employ them.

\begin{figure}[t]
\centering\includegraphics[width=\columnwidth]{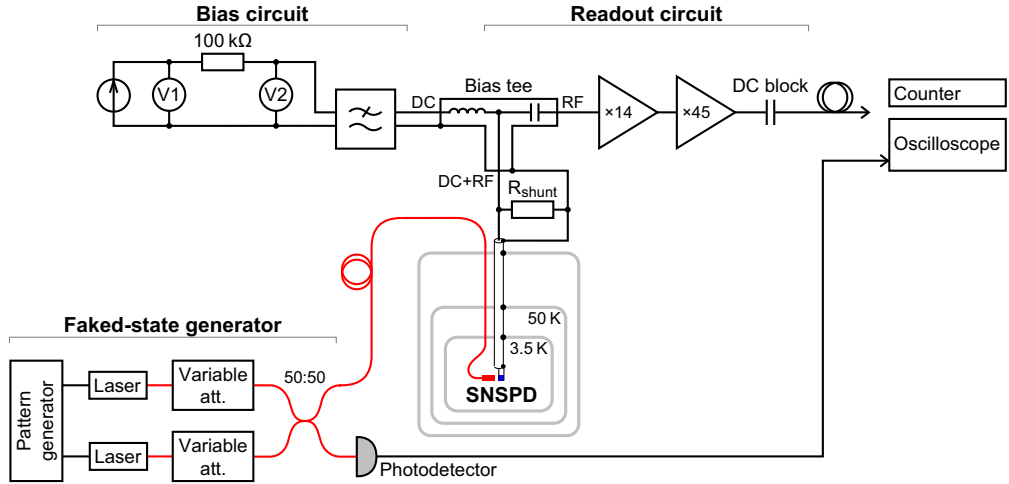}
\caption{\label{fig:setup} Experimental setup. SNSPD is cooled to $3.5\,\kelvin$ in a closed-cycle refrigerator, custom designed around a Cryomech PT-403 pulse tube cold head, and connected to room-temperature electronics via $\sim 1\,\meter$ long $50\,\ohm$ coaxial cable. The electronics consists of the bias and readout circuits. The bias circuit is composed of a battery-powered low-noise voltage source (Stanford Research Systems SIM928), $100\,\kilo\ohm$ resistor converting voltage into bias current, two voltmeters (Stanford Research Systems SIM970), and low-pass filter (Mini-Circuits BLP-1.9+). The bias current is applied to the SNSPD via the direct-current (DC) port of a bias tee (Picosecond Pulse Labs 5575A-104, $10\,\kilo\hertz$--$12\,\giga\hertz$ radio-frequency (RF) port bandwidth). The readout circuit uses two radio-frequency amplifiers (RF~Bay LNA-580, $23\,\deci\bel$ gain $10$--$580\,\mega\hertz$ bandwidth, and RF~Bay LNA-1000, $33\,\deci\bel$ gain $10\,\mega\hertz$--$1\,\giga\hertz$ bandwidth), and a DC block (Mini-Circuits BLK-18-S+, $10\,\mega\hertz$--$18\,\giga\hertz$ bandwidth). The pulses are registered by either a counter (Agilent 53131A) or an oscilloscope (Agilent Infiniium DSO80804A, $8\,\giga\hertz$ $40\,\giga\text{samples}\per\second$). The SNSPD is illuminated via a single-mode fiber (Corning SMF28e) shown in red, by light formed by the faked-state generator. The latter consists of a pulse pattern generator (Agilent 81110A), two $1550\,\nano\meter$ semiconductor laser diodes (one Thorlabs LPS-1550-FC and one Thorlabs LPSC-1550-FC), two optical variable attenuators (Hewlett-Packard 8156A) and a 50:50 ratio fiber beamsplitter providing two identical optical outputs. One output is connected to the SNSPD, while the other is monitored with a classical photodetector (Thorlabs DET01CFC, DC--$1.2\,\giga\hertz$ bandwidth).}
\end{figure}

\section{\label{sec:experiment}Experiment}

We have tested five SNSPD devices, summarised in Table~\ref{tab:devices}. The majority of data presented here was obtained from device~1. This detector is of the superconducting nanowire avalanche photon detector (SNAP) type with sections of nanowires connected in parallel \cite{ejrnaes2007,marsiliNanoLett2011,tanner2012,Heath2014,Ejrnaes2009}. This configuration is advantageous for reducing nanowire dimensions, in order to increase device efficiency while maintaining usable current levels in the detector and for reducing reset times for achieving higher count rates. This detector implementation is likely to be used in future high speed and detector efficiency QKD systems. However for completeness a representative range of detector types were tested. These included traditional meander-patterned devices on a variety of substrates, such as those used in several practical demonstrations of QKD (device~3) \cite{sasaki2011,clarke2011}. Next-generation optical cavity enhanced detectors were included as well (devices~4 \&~5) \cite{tanner2010}, which are now becoming available for QKD implementations. The same blinding attack technique was successful with all detector types.

Our experimental setup (Fig.~\ref{fig:setup}) represents a typical detector configuration used in QKD experiments \cite{hadfield2006}. The SNSPD device is biased at about $0.9$ of its critical current (specific device properties such as critical current at the operating temperature are listed in Table~\ref{tab:devices}). The bias is applied by a battery-powered low-noise current source connected via a bias tee. The important feature of the scheme is the presence of a shunt resistor $R_{\text{shunt}}$ that prevents latching (typically a $50\,\ohm$ resistor) \cite{hadfield2005}. This resistor creates a low-impedance mismatch point $\sim 1\,\meter$ away from the SNSPD along the $50\,\ohm$ coaxial radio-frequency (RF) cable. A reverse-polarity pulse reflected from this impedance mismatch reaches the SNSPD about $10\,\nano\second$ after hotspot formation, and lowers the voltage across the device. If the hotspot has failed to dissipate and persists by Joule self-heating, this reflected pulse can remove electrical power from it aiding reset from the latched state. On longer timescales it is important that the shunt resistance is much lower than the SNSPD hotspot resistance. The shunt resistor provides an alternate current path, reducing current flow through the detector allowing cooling back to the superconducting state. The latter recovery mechanism is of particular importance to the detector blinding attack described in this paper. In this circuit configuration, SNSPD can be reliably operated at a higher bias current and higher photon detection efficiency than in the configuration without $R_{\text{shunt}}$. The pulse readout circuit consists of AC-coupled amplifiers with combined gain of $56\,\deci\bel$ and $10$--$580\,\mega\hertz$ frequency range. The detector output signal is observed with an electronic counter and an oscilloscope. The SNSPD is illuminated via single-mode fiber connected to the output of a faked-state generator. The faked-state generator allows the formation of arbitrary illumination diagrams with two distinct optical power levels at the SNSPD, in addition to zero power level. This is achieved with a pulse pattern generator powering two $1550\,\nano\meter$ laser diodes, followed by optical variable attenuators to set the power levels. The output of the faked-state generator simulates illumination diagrams that the SNSPD would receive if it were a part of a QKD system under attack \cite{lydersen2011c}.

\begin{figure}[t]
\centering\includegraphics{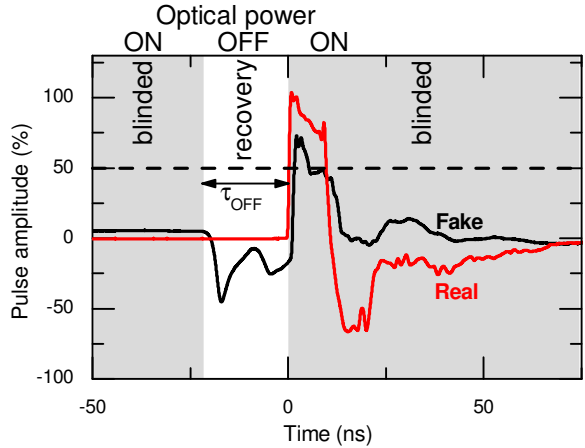}
\caption{\label{fig:pulses} Output pulses from the experimental setup described in Fig.~\ref{fig:setup} under normal single-photon illumination (red trace) and under bright-light illumination manipulating the detector (black trace). Laser illumination for the blinded case is illustrated by the plot shading. Time $t = 0\,\nano\second$ is the point at which the fake output pulse is triggered.}
\end{figure}

A typical output pulse from this setup is shown in Fig.~\ref{fig:pulses}, triggered by the incidence of a single photon. The normal character of an SNSPD output pulse includes a sharp leading edge as the detector becomes resistive and the current is forced out, followed by a slower recovery as the current returns to the detector. The shape of the observed recovery signal is highly dependent on amplifier bandwidths and reflections from components (such as the shunt resistor used in this setup). Note that this oscilloscope trace is not an accurate representation of the current flow returning to the device. The critical part of the pulse is the sharp clean leading edge on which counting electronics is normally triggered, providing the advantageous timing properties of SNSPDs. The observed leading edge is also dependent on amplifier bandwidth and hotspot resistance \cite{OConnor2011}. Hotspot growth time (typically $< 100\,\pico\second$ \cite{yang2007,marsili2011}) is normally short in comparison to the observed pulse rise time. In our experimental setup, the latter is limited by the first $580\,\mega\hertz$ bandwidth amplifier (Fig.~\ref{fig:setup}).

Lydersen {\it et al.}\ considered artificially generating pulses in SNSPDs through two methods \cite{lydersen2011c}. The first involved latching the detector into the resistive state, through a short bright-light illumination, from which the detector does not recover. Fake detector pulses were generated through subsequent bright pulses causing variation of the device resistance. However, this attack is effectively defeated by the inclusion of a shunt resistor \cite{hadfield2005} (or other reset circuit) as implemented in our standard experimental setup (Fig.~\ref{fig:setup}), and also in some QKD demonstrations \cite{hadfield2006} in order to allow stable long-term detector operation. In this paper we describe the extension of the second method put forward by Lydersen {\it et al.}\ of blinding the detectors to incoming single photons through continuous bright-light illumination (of the order of 1 to 100 \micro W in this study depending on individual SNSPD characteristics). We find that with careful control it is possible to generate fake detector output signals reliably on-demand with timing properties better than in the single-photon case.

\section{\label{sec:detector-control}Detector control}

\subsection{\label{sec:qkd-schemes}Applicability to different QKD schemes}

Since our testing was performed on a stand-alone detector, we need to briefly address the question how this applies to hacking a complete QKD system. In a detector control attack, Eve performs an intercept-resend in the transmission line. She uses a replica of Bob's setup to detect all quantum states emitted from Alice, then generates and sends bright-light faked states to Bob that attempt to replicate Eve's detection results in Bob's detectors \cite{makarov2005,makarov2009}. Note that Bob has two or more detectors. The simplest version of this attack requires that Eve can specify deterministically which detector in Bob clicks and when it clicks, at her will. To do so she needs to form bright-light pattern at the target detector that causes it to click with $100\%$ probability, while all other detectors in Bob receive bright-light pattern that keeps them silent. The hacking method employed by Eve, and the sucess thereof, depends on the optical layout in Bob. For the purpose of the following analysis, we can broadly classify Bob's optical schemes into three categories.

The first category contains passive measurement schemes in which Eve can, by choosing appropriate polarization or phase of bright light, reduce light power at a selected detector by $20\,\deci\bel$ (100 times) or more for an arbitrary time period, while keeping the other detectors illuminated \cite{makarov2009}. Examples of such schemes are passive-basis-choice BB84 protocol and DPS-QKD protocol. The $20\,\deci\bel$ figure is a typical extinction ratio of Bob's polarization- or phase-selective component, such as a polarizing beamsplitter or Mach-Zehnder interferometer. While the power at the selected detector is reduced greatly, other detectors in Bob may receive excess power in the form of a surge of up to $3\,\deci\bel$ (a twofold increase in power) \cite{makarov2009,lydersen2011c}. Most QKD schemes so far tested with SNSPDs are of this type \cite{collins2007,takesue2007,Wang2012,liu2010,Honjo2008,Choi2010,clarke2011}. We thus base our stand-alone detector testing on a control diagram that alternates between $20\,\deci\bel$ optical power drop and $3\,\deci\bel$ surge around a steady blinding power level, as will be detailed in Section \ref{sec:pulse-generation}.

A second smaller category contains schemes where Bob uses a randomly-driven modulator for active basis choice in BB84 protocol \cite{hadfield2006,rosenberg2009}. In these schemes Eve can drop power at the target detector by $20\,\deci\bel$ conditional on a specific basis choice by Bob, while for other basis choices the power at all detectors will stay around the blinding level \cite{makarov2009,lydersen2010a}. The time duration for which Bob keeps one choice of basis applied to his modulator presents an additional constraint on Eve. If this time is greater than the time duration for which Eve needs to reduce the power delivered to the target detector, then her attack is in essence equivalent to the previous category. However in high-speed QKD implementations this condition may not hold and Eve's life becomes more complicated. In the latter case the devil is in the details: Eve may or may not be able to hack, depending on the exact particulars of the technical implementation. We do not consider the latter case, because there is no stable reference QKD implementation that we could obtain and examine in detail. Unfortunately all QKD implementations using SNSPDs have to date been laboratory prototypes -- no complete commercial system yet exists.

A third, also smaller, category contains certain time-bin encoding schemes in which one or more detectors are essentially connected straight to the communication line and are not selective on any light parameter. These include implementations of COW protocol \cite{stucki2009} and time-bin-encoded BB84 protocol \cite{Tanaka2008}. In these schemes Eve will again be constrained by the exact details of the technical implementation (among other details, by the splitting ratio of the asymmetric beamsplitter \cite{lydersen2011}), and will have to analyse detector behavior beyond the level tested in this paper.

Looking beyond the optical layout in Bob, the readout circuitry can also affect Eve's attack strategy. The discriminator circuits used to convert the analog pulse shown in Fig.~\ref{fig:pulses} into a digital detection signal vary, and are never described in sufficient detail in papers about QKD implementations. For example, it is usually not stated at what level the discriminator threshold is set. SNSPD pulse shape has the steepest and cleanest gradient mid-way through the leading edge, making $50\%$ of the output peak height a sensible robust choice of discriminator level for-low jitter QKD operation. We thus assume in our testing that Bob uses a fast constant-level discriminator with threshold set at $50\%$ of the single-photon pulse height. As will be obvious below, the attack works in a range of discriminator levels centered around $50\%$, and Section \ref{sec:pulse-character} discusses how fake pulse height can be varied during attack optimisation.

\subsection{\label{sec:pulse-generation}On-demand fake pulse generation}

When illuminated with a bright `blinding' pulse of light, the detector becomes resistive over a larger area than the single hotspot generated by single photon absorption. In the single-photon detection case the resistive region, or hotspot, grows due to Joule heating with the energy dependent upon $I^2 L$, where $I$ is the bias current and $L$ is the kinetic inductance of the detector. Once the bias current is shunted out from the detector, the hotspot dissipates on a time scale determined by the rethermalisation of the nanowire with the substrate. This mechanism has been modelled in detail by others \cite{yang2007,marsili2011}.

In the bright-light case, we suggest that the resistive region is maintained through the direct absorption of the incident laser power in excess of the rethermalisation or cooling power of the SNSPD environment. It is interesting to note that very approximately, given bias current of the order of $10\,\micro\ampere$, kinetic inductance of $1\,\micro\henry$ and hotspot formation time of $100\,\pico\second$, the power dissipated during hotspot formation is $\sim 1\,\micro\watt$. This power is sufficient to cause the hotspot to grow rather than rethermalise. This agrees well with the blinding powers required to maintain the devices in a resistive state ($> 1\,\micro\watt$ or $-30\,\textrm{dBm}$, see Table~\ref{tab:devices}).

\begin{figure}[t]
\centering\includegraphics[width=87mm]{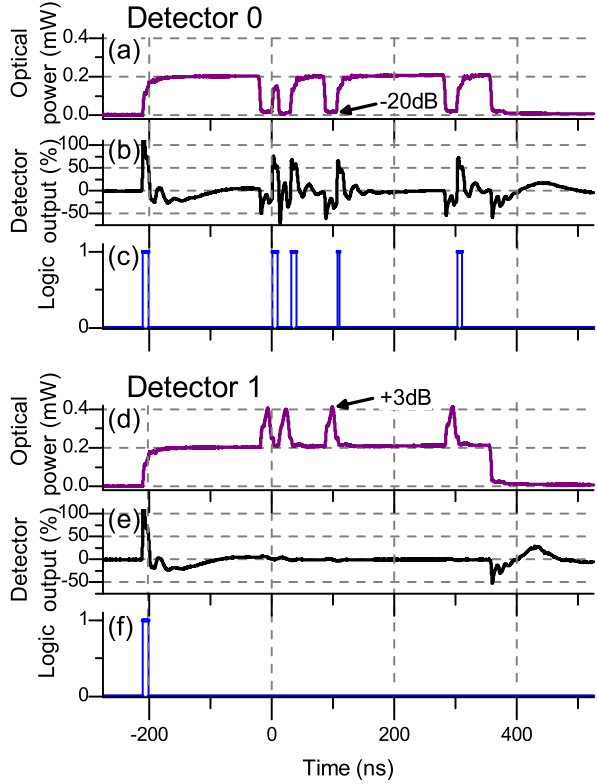}
\caption{\label{fig:pulse-trains} Simulated control diagrams two detectors would receive inside Bob whose scheme allows to redistribute optical power between the detectors (i.e.,\ measurement scheme of the first category in Section \ref{sec:qkd-schemes}). Detectors are controlled through blinding with a bright laser pulse at time $t = -200\,\nano\second$, followed by variation of the blinding laser power by $-20\,\deci\bel$ on detector~0, and corresponding variation of $+3\,\deci\bel$ on detector~1. Optical power at both detectors during the attack is shown in oscilloscope traces (a) and (d), while analog detector outputs are shown in oscilloscope traces (b) and (e). Corresponding logic pulses obtained by passing the analog signal through a $50\%$ fixed-threshold discriminator are shown in (c) and (f).}
\end{figure}

Under blinding illumination, current is diverted from the detector causing an output pulse [see pulse at $t \sim -200\,\nano\second$ in Fig.~\ref{fig:pulse-trains}(b) and ~\ref{fig:pulse-trains}(e)]. If the bright illumination continues, the detector remains in the resistive state and is no longer sensitive to incident photons. However, if the bright illumination is stopped (or its power is decreased sufficiently, $20\,\deci\bel$ attenuation is shown to be sufficient in Fig.~\ref{fig:pulse-trains}) for a short period of time (e.g.,\ $< 50\,\nano\second$), the nanowire rethermalises. It then once more becomes superconducting, and the current starts to return to the detector at a rate defined by the superconducting kinetic inductance of the SNSPD $L$ and the circuit resistance. Recovery of the SNSPD after the blinding attack is somewhat different than recovery from single photon detection. Excess laser power has been absorbed into the detector, driving a large area resistive and causing a local rise in temperature. The need to rethermalise in addition to the normal return of current to the SNSPD extends recovery timescales dependent on the excess blinding energy deposited (or timescale of the attack). If enough of the bias current was allowed to return to the detector, it would once more become single-photon sensitive (after time $\tau_{\text{recovery}}$), and would also exhibit dark counts. Note that it does not require the full bias current to have returned to the nanowire before the detector can exhibit a photoresponse or produce dark counts \cite{marsili2012,burenkov2013}. However, if the bright illumination is re-applied before this time, after $\tau_{\text{OFF}} < \tau_{\text{recovery}}$ optimised experimentally in this work, the proportion of the current that had already returned to the detector is again forced out as the nanowire returns to the resistive state. This elicits another controlled fake output pulse from the detector while maintaining the SNSPD in a `blinded' state. An example of this fake pulse is shown in Fig.~\ref{fig:pulses}. This is the basis of the detector attack described in this paper.

In the manner described above, an attacker can blind an SNSPD and elicit `fake' output pulses on-demand. This is shown explicitly in the top half of Fig.~\ref{fig:pulse-trains}. An initial output pulse occurs when the blinding illumination is initiated at $t \sim -200\,\nano\second$, and subsequent controlled pulses are generated on-demand through brief reductions in the blinding illumination for time $\tau_{\text{OFF}}$ $<$ $\tau_{\text{recovery}}$ (in this case $\tau_{\text{OFF}}$ = $20\,\nano\second$). Once the control diagram was optimised, a fake output pulse occurred on every observed attempt in the authors' experiment with apparent 100\% probability.

In order to achieve successful manipulation of the variety of SNSPDs tested in this work, some variation of parameters was needed, primarily blinding power and $\tau_{\text{recovery}}$ (see Table~\ref{tab:devices}). Devices were biased and operated as the authors would normally use them in experiments; only blinding attack parameters were optimised. While new generations of SNSPDs with near-unity detection efficiency using new materials such as tungsten silicide \cite{marsili2012} were not tested in this study, their operating principle is the same. Variations in thermal and electrical properties are likely to require similar optimisation of the blinding parameters, but no change in the principle of the attack. In practice it may seem impractical to determine the correct blinding parameters to attack a system. However we assume, in accordance with Kerckhoffs' principle \cite{kerckhoffs1883} (a cryptosystem should be secure even if everything about the system except the key is public knowledge), that the attacker knows all details of the devices, settings and protocols used. In practice, as detectors and commercial QKD systems develop, it is likely the detectors will have highly repeatable characteristics. It then becomes realistic to fully dismantle and analyse a sample of a commercial product to obtain starting values of these parameters in advance of attacking a QKD implementation. The attack may then begin to be applied intermittently while analysing the public communication between Alice and Bob and fine-tuning attack parameters \cite{makarov2005,gerhardt2011}. Eve would subsequently switch over to continuous attack with real-time adjustment of parameters, if necessary.

\subsection{\label{sec:pulse-character}Pulse and recovery characteristics}

The characteristics of the fake pulse seen in Fig.~\ref{fig:pulses} are qualitatively similar to those of the real pulse: a sharp leading edge followed by a slow recovery. Amplitude of the fake pulse is reduced, because only a fraction of the full device current has returned to the detector before the fake pulse is triggered. If a longer pause is left before resuming the full blinding laser power, the fake pulse amplitude is increased. However, with pauses of duration closely approaching $\tau_{\text{recovery}}$, there is a finite probability of a count occurring during the recovery from the blinded state, which is undesirable for full detector control. For the fake pulse outputs demonstrated in this paper, $\tau_{\text{OFF}}$ was kept sufficiently below $\tau_{\text{recovery}}$ (in this case $\tau_{\text{OFF}}$ = $20\,\nano\second$). Then counts due to recovery from the blinded state did not occur during the attack, instead the fake pulse was generated returning the detector to the blinded state. This was confirmed in the good jitter characteristics of the fake pulses, discussed in Section \ref{sec:jitter}. Fake pulse amplitude can be increased at the cost of a finite probability of a detector pulse occurring before the intended fake pulse.

\begin{figure}[t]
\centering\includegraphics{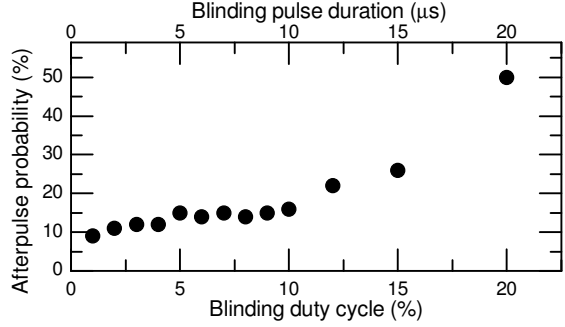}
\caption{\label{fig:Afterpulsing} Probability of an afterpulse occurring when the blinding pulse is stopped, dependent on the fraction of the time the detector is illuminated by the blinding pulse. Blinding attack repetition rate was kept constant at $10\,\kilo\hertz$, while blinding pulse duration was varied. Since many fake pulses would be generated during each blinding cycle, fake detection rate would be much higher than $10\,\kilo\hertz$.}
\end{figure}

However, counts during recovery from the blinded state are common if the blinding attack is stopped (e.g.,\ $t > 400\,\nano\second$ in Fig.~\ref{fig:pulse-trains}), occurring with a probability $10$--$16\%$ when the detector is blinded for $1$--$10\,\micro\second$, see Fig.~\ref{fig:Afterpulsing}. The recovery of the detector from the blinded state is different from normal single-photon detection recovery (which can also stimulate afterpulsing \cite{burenkov2013}), as in the blinded case the detector must rethermalise to the base temperature before the system fully returns to normal operation. If carefully applied, the blinding power need not heat the detector excessively and the thermalisation time only slightly extends the recovery, which is still dominated by the current return time to the detector. However the dynamics of this recovery are affected by the temperature from which the SNSPD is rethermalising, hence the dependence of afterpulsing probability on blinding duty cycle as in Fig.~\ref{fig:Afterpulsing}. An additional contribution to afterpulses may be single-photon detection of photons delayed in the optical scheme via multiple back-and-forth reflections of the bright blinding pulse. The observed output signal during recovery from the blinded state is most clearly seen in Fig.~\ref{fig:pulse-trains}(e) after $t = 350\,\nano\second$.

The presence of afterpulses should not stop the attack. In Fig.~\ref{fig:pulse-trains}(b) at $t = 0\,\nano\second$ two fake pulses are generated at a repetition period of $30\,\nano\second$. After the first pulse, $10\,\nano\second$ of bright light is required to return the detector to the blinded state before a second fake pulse can be generated with $\tau_{\text{OFF}}$ = $20\,\nano\second$. In this manner, fake pulses can be generated at a repetition rate of $33\,\mega\hertz$. While these parameters vary between detectors (see last row in Table~\ref{tab:devices}), by the very nature of the attack discussed above $\tau_{\text{OFF}}$ is kept well below $\tau_{\text{recovery}}$ (in this case at 50\%). In normal QKD operation, the maximum single-photon detection rate would be 1/$\tau_{\text{recovery}}$ \emph{with a unity efficiency detector.} The hacker can match or better this rate, with significant further gains available when compared to a non-unity efficiency single photon detector in Bob.

For the detector studied in detail in this paper, if kept under Eve's control for $10\,\micro\second$ every $100\,\micro\second$, an afterpulse will occur 16\% of the time (Fig.~\ref{fig:Afterpulsing}). In each $10\,\micro\second$ cycle the hacker can generate 333 fake pulses, or if matching the maximum single photon detection rate only 250 fake pulses, with an average of 0.16 afterpulses. As such, afterpulses per fake pulse occur at a rate of $0.06\%$, adding only a small contribution to the error rate.

\subsection{\label{sec:jitter}Jitter}

\begin{figure}[b]
\centering\includegraphics{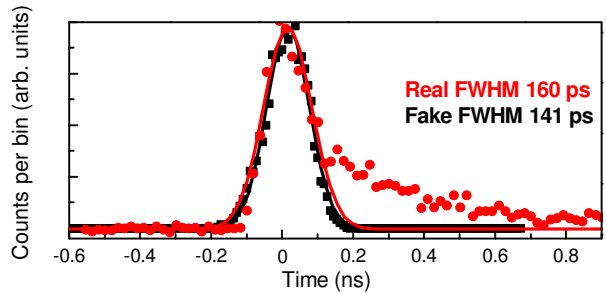}
\caption{\label{fig:jitter} Comparison of timing jitter measured in the experimental setup described in Fig.~\ref{fig:setup}, for device~1. Timing distribution due to single-photon illumination (red circles) and manipulation through bright-light illumination (black squares) is shown, together with Gaussian fits. FWHM time widths are $160$ and $141\,\pico\second$, respectively. Jitter is measured at a fixed threshold level set at 50\% of the amplitude of a single-photon detection pulse.}
\end{figure}

For good detector control, the timing jitter of the fake electrical output pulses must be comparable or better than that of the real response. This is shown in Fig.~\ref{fig:jitter}. As long as the pause in the blinding pulse is kept below $\tau_{\text{recovery}}$, the jitter achieved is as good or better than for single-photon response, for all detectors tested. While normal SNSPDs suffer from some variation in timing response over the detector area due to varying hotspot resistance of $\sim 1\,\kilo\ohm$ \cite{OConnor2011}, in the case of the blinding attack the SNSPD switches to a very high resistance every time, giving a sharper leading edge to the pulse and improved timing jitter.

Additionally, for the real single-photon case shown in Fig.~\ref{fig:jitter} a tail is observed on the jitter histogram, characteristic of the avalanche process for parallel wire (SNAP) detectors \cite{ejrnaes2007,marsiliNanoLett2011,tanner2012,Heath2014,Ejrnaes2009} (devices 1 \& 2). The tail is not observed for the standard meander SNSPDs (devices 3--5). This tail is not present in the fake jitter histogram for any of the five devices, as the higher power of the blinding pulse ensures immediate cascade of the detector into the resistive state. Improved FWHM characteristics of the faked detector response for all detectors tested offer Eve some leeway in her hacking attack: for example, improved error rate here may be used to compensate for any increased errors due to afterpulses.

\subsection{\label{sec:detector-control-summary}Summary}

Experimental results obtained with simulated control diagrams shown in Fig.~\ref{fig:pulse-trains} show that the detectors are successfully manipulated with only $20\,\deci\bel$ variation in blinding power. These diagrams should be fully reproducible when attacking the majority of QKD schemes using SNSPDs \cite{collins2007,takesue2007,Wang2012,liu2010,Honjo2008,Choi2010,clarke2011}, making these schemes vulnerable to the detector control attack. The control diagrams can be adapted to minor variations between individual detectors in the QKD system, such as different values of $\tau_{\text{recovery}}$ and minimum blinding power. For the remaining QKD schemes \cite{hadfield2006,rosenberg2009,stucki2009,Tanaka2008} these control diagrams may not directly apply, yet should be good starting point for refining the attack. For the latter schemes, detailed investigation of the implementation details and a tailored advanced attack tactics would be required. For example, for an active basis choice scheme that switches measurement bases faster than $\tau_{\text{recovery}}$, Eve could try to send faked states tailored to certain sequences of bases. We did not investigate these schemes owing to the lack of a stable reference implementation such as a commercial QKD system that uses SNSPDs.

\section{\label{sec:countermeasures}Countermeasures}

An attack such as that described in this paper will always be dependent on the exact configuration of the QKD system. This paper attempts to demonstrate that vulnerability to attack exists in stand-alone SNSPDs of all configurations available to the authors, with only minor adjustment of parameters (see last two rows in Table~\ref{tab:devices}). A further investigation would have to target a complete QKD system containing SNSPDs. This would be a level of effort outside the scope of this paper, especially as no commercial QKD systems using SNSPDs are yet available as a benchmark. However, it is worth considering countermeasures that may remove this vulnerability in the future.

There are two main forms of countermeasure available to eliminate the security loophole demonstrated here. The preferred action is to include the equipment imperfections in the security model, as for example is done in the measurement-device-independent QKD scheme \cite{lo2012,braunstein2012,rubenok2012} where the detector system is moved outside of the security proof. However, in practice, patches to rule out already demonstrated attacks on existing systems are often considered first, while not offering any guarantee that the vulnerability can be eliminated \cite{yuan2011,lydersen2011d,yuan2011a,legre2010}. Below we give some ideas for these latter kind of `band-aid' countermeasures applicable to the attack described here.

As mentioned in section 2, the inclusion of $R_{\text{shunt}}$ in the experimental design shown in Fig.~\ref{fig:setup} offers a countermeasure to the first attack described by Lydersen {\it et al.} \cite{lydersen2011c}. Alternate reset circuits including systems which actively reset the bias to avoid detector latching may have significant implications for the operation of the attack and may form the basis of a countermeasure. It may be that such active reset circuits are once again vulnerable to the first type of attack described by Lydersen {\it et al.} \cite{lydersen2011c}. Due to lack of availability to the authors, such active reset circuits have not been investigated in this work.

For the passive reset circuit described here, a countermeasure uses the feature that if the detector under blinding attack from bright-light illumination is under DC electrical monitoring, a small increase in the average resistance can be observed, limited by $R_{\text{shunt}}$. This manifests itself as a measurable average voltage drop across the DC bias port (measured by voltmeter V2 in Fig.~\ref{fig:setup}), dependent on the duty cycle of the blinding attack. The reading on V2 increased linearly from $0.2\,\milli\volt$ to $0.5\,\milli\volt$ with blinding duty cycle varying from 0 to 50\%. This is at the limit of the resolution of the standard voltmeter used here. The fractional change in measured resistance was slight in this demonstration especially at short blinding pulse duration (or blinding duty cycle). It can be imagined that more sensitive device monitoring of the correct bandwidth may enable easier detection of attacks that put the detector into a resistive state for a greater time than expected in normal operation. However, it should be noted that in high bit rate QKD the detector will be running at close to its maximum count rate. After each count, during detector recovery, a finite resistance would also be measured on V2. The wise hacker injecting high bit rate fake detector pulses will be aware of this and may be able to keep the blinding duty cycle low, keeping variation on V2 comparable to that caused by high bit rate QKD. It can be imagined that attacks may be limited to short periods of detector blinding.

A further countermeasure could be implemented as follows: The shape of the fake output pulses in this attack are highly dependent on the amplifiers used in the system. The setup used here is the standard arrangement employed in the majority of the authors' work. Real and fake pulses demonstrated here have the same important features (see Fig.~\ref{fig:pulses}), suitable for triggering a discriminator in a QKD system. However, we also tested other configurations of amplifiers. The leading edge of the fake pulse is maintained with the range of amplifiers tested. However, if DC-coupled amplifiers are used (instead of the AC-coupled standard amplifier chain in Fig.~\ref{fig:setup} that has $10\,\mega\hertz$ low frequency cut-off), a different characteristic is seen. While the detector is in the blinded state, a constant output level is observed, only relaxing during the recovery phase. A pulse is still observed when the detector is switched to the blinding state, but the recovery does not match that of a real pulse. This would still be suitable for triggering many types of discriminators. The situation may be further complicated by the use of cryogenic amplifiers, or complete superconducting readout circuits \cite{Terai2010} currently being developed. If a fast analog-to-digital converter (ADC) is used after the analog electronics instead of a simple threshold discriminator, it may be possible to distinguish real and fake pulses via more detailed automated analysis of the signal shape.

Further countermeasures of this type may also be possible, trying to distinguish distorted pulse shape and other abnormalities. However the authors believe that the type of attack described here is less dependent on the precise electrical circuit than the latched-state attack originally described by Lydersen {\it et al.} \cite{lydersen2011c}, and could be developed further by potential hackers in response to simple countermeasures.

\section{\label{sec:conclusion}Conclusion}

In this paper, we have demonstrated further vulnerabilities in SNSPDs used in QKD systems. The purpose of this study has been to demonstrate the detailed operation of an attack on SNSPDs and moreover to consider the weaknesses of such an attack.  Although hacking of a real QKD system has not been demonstrated, this study nevertheless provides an important signposting for future QKD system development. This bright-light blinding and control has been successfully demonstrated on a range of currently available SNSPD devices of different types on a variety of substrates. The attack has been shown to produce fake pulses and pulse trains on-demand with timing characteristics better than the detector's normal response. It has been shown that for many QKD schemes, multiple detectors in Bob could be controlled individually. As such, it is suggested that careful consideration of the full QKD system and security model including detectors is needed in the development of commercial apparatus, before robust security claims can be made.

\section*{Acknowledgments}
The authors thank L.\ San Emerito Alverez, W.\ Jiang and Z.\ H.\ Barber at the University of Cambridge, UK, S.\ Miki, Z.\ Wang and M.\ Sasaki at NICT, Japan, and S.\ N.\ Dorenbos and V.\ Zwiller at TU Delft, Netherlands for supplying devices.
V.M.\ thanks Industry Canada for support.
R.H.H.\ acknowledges a Royal Society University Research Fellowship.
R.H.H.\ and M.G.T.\ acknowledge support from the UK Engineering and Physical Sciences Research Council (EPSRC grants EP/F048041/1, EP/G022151/1 and EP/J007544/1).

%

\end{document}